\documentclass[aps,12pt]{revtex4} 
\usepackage{graphicx}				
\usepackage{amssymb}

\newcommand{\R}{\mathbb R}

\begin{document}

\title{The world is discrete.}

\author{Olaf Dreyer}
\email{olaf.dreyer@gmail.com}
\affiliation{Dipartimento di Fisica and INFN,\\ ``Sapienza" University of Rome,\\ P.le A. Moro 2, 00185 Roma, EU }

\begin{abstract}
We argue that the scale-free spectrum that is observed in the cosmic microwave background is the result of a phase transition in the early universe.  The observed tilt of the spectrum, which has been measured to be $0.04$, is shown to be equal to the anomalous scaling dimension of the correlation function.  The phase transition replaces inflation as the mechanism that produces this spectrum.  The tilt further indicates that there is a fundamental small length scale in nature that we have not yet observed in any other way.
\end{abstract}

\maketitle

\section{Introduction}
Observations of the cosmic microwave background find that the primordial spectrum of the metric perturbations is given by\cite{planck}
\begin{equation}\label{eqn:spectrum}
\delta^2_\Phi \sim k^{-0.04}.
\end{equation}
The most widely studied theory that explains this spectrum is inflation (see \cite{slava}.  For a different point of view see \cite{cyclic}).  Inflation accounts for this spectrum by introducing a new length scale into the problem:  the Hubble radius of the inflating universe.  When the wave length of a mode of the inflaton becomes larger than the Hubble scale the evolution of the mode effectively stops.  The amplitude of the field at the moment of horizon crossing determines the amplitude of the perturbation.  Because inflation has to end the Hubble scale changes at the end of inflation.  It is this changing Hubble scale that is responsible for the tilt in the above spectrum.  To achieve this exit a particular potential for the inflaton has to be postulated. 

Instead of postulating a new particle and a very special potential for it we propose that a phase transition in the early universe is the reason for the spectrum of equation (\ref{eqn:spectrum}).  Phase transitions in the early universe are nothing new.  It is widely believed that we arrived at the current gauge group of the standard model by a series of symmetry breaking processes.  Unlike the transition that we are thinking about here though these processes happened in ordinary spacetime.  We are instead talking about a more radical kind of phase transition;  namely a transition in which space itself emerged.  On this side of the transition we have our well known notions of space and elementary particles whereas on the other side of the transition space and particles are completely different or do not exist at all.

Let us describe the process that we have in mind.  We start with the system (the universe) in its unordered phase at a temperature $T$ above a critical temperature $T_c$.  As we approach the transition temperature the correlation length $\xi$ and the relaxation time $\tau$ both grow.  If we define
\begin{equation}
\epsilon = \frac{T-T_c}{T_c}
\end{equation}
then the behavior of $\xi$ and $\tau$ near the transition is given by \cite{chaikin,Kibble:fm,Zurek:2008ex}
\begin{eqnarray}
\xi & \sim & \vert t \vert^{-\nu} \\
\tau & \sim & \vert t \vert^{-\nu z},
\end{eqnarray}
for some critical exponents $\nu$ and $z$.  At some point the rate of change $\dot\epsilon/\epsilon$ will exceed the inverse relaxation time $\tau^{-1}$.  At this point the evolution of the system freezes.  Let us call the correlation length of the system at this point $\xi_0$.  The evolution of the system will not continue again until we have passed the transition temperature $T_c$ and the rate of change is again smaller than the inverse relaxation time in the new phase \cite{Kibble:fm,Zurek:2008ex}.  What happens now is the crucial part in the derivation of the spectrum.  The system will relax to a ground state in the correlated domains of size $\xi_0$.  If the transition is a symmetry breaking transition this is the moment the symmetry brakes.  We are then left with domains of size $\xi_0$ that each represent a different ground state.  The spectrum that we are after is the result of this transition to the new ground state.  Before the transition we have a distribution of regions of all orientations and sizes up to $\xi_0$.  In the transition some of these smaller regions have to flip to form the new ground state.  This flipping of the smaller regions will leave behind perturbations in the newly formed ground state that carry the signature of the state of the system just before the transition.  If $\theta$ is an order parameter than we want to calculate the spectrum $\delta_\theta^2$ of the order parameter just before the transition.  We will show that this spectrum is naturally scale free:
\begin{equation}
\delta^2_\theta \sim k^0
\end{equation}
In turns out that this is not the whole story.  This would be the spectrum if there would be no other small length scale in the problem.  In solid state physics the atomic spacing provides another length scale that slightly changes the exponent.  We face a similar situation here with the difference that the constituents are not the atoms that we know.  The critical exponent that is relevant here is the anomalous dimension $\eta$ of the correlation function
\begin{equation}
G(r) \sim r^{-d+2-\eta}.
\end{equation}
The typical value for $\eta$ is $0.04$ (see table \ref{tab:eta}).  We will argue that the spectrum of $\theta$ is not flat but that instead
\begin{equation}
\delta^2_\theta \sim k^{-\eta}.
\end{equation}
This is exactly the spectrum of the metric perturbations of equations (\ref{eqn:spectrum}).  This is our central result.

There are now three attitudes one can take towards this result.  One possibility is to say that this just represents a numerical coincidence and does not tell us anything significant about early universe cosmology.  Another possibility is to say that this picture provides an effective description of what is going on.  What needs to be done then is to find the fundamental theory whose degrees of freedom are effectively described by the order parameter $\theta$.  The last possibility is to take this result literally and equate the order parameter $\theta$ with the metric perturbation $\Phi$.  This is the path that was advocated in \cite{dreyer}.  The cosmological microwave background then becomes the first experimental hint that tells us where to look for a quantum theory of gravity.  We infer that gravity is emergent,  that there is a fundamental discreteness,  and that the non-perturbative degrees of freedom of the vacuum have to be identified with gravity.

\section{The spectrum}
For concreteness we assume that we are dealing with the simple case of a second order phase transition with a scalar  order parameter $\theta$.  We want to calculate the spectrum of the order parameter just before the domains of size $\xi_0$ are formed.  For the domains of size $\xi_0$ to form the regions of smaller size inside the domain have to flip to form the new ground state.  The spectrum is given by  
\begin{equation}
\delta^2_\theta = k^3 \lim_{V\rightarrow\infty} \frac{\vert\theta_V(k)\vert^2}{V}
\end{equation}
Here $\theta_V(k)$ is the Fourier transform of $\theta$ restricted to the volume $V$.  The order parameter $\theta(x)$ is the sum of the contributions from the different regions.  If $I$ is the set of regions then
\begin{equation}
\theta(x) = \sum_{i\in I} \theta^i(x).
\end{equation}
Because we are interested in the ensemble average we find
\begin{equation}
\vert\theta_V(k)\vert^2 = \sum_{i\in I} \vert\theta^i_V(k)\vert^2.
\end{equation}
The contribution from a region $i\in I$ only depends on its size.  Let us define $\nu(l)$ so that
\begin{equation}
V \nu(l) dl
\end{equation}
is the number of regions with sizes between $l$ and $l+dl$ in the volume $V$.  Then
\begin{equation}
\vert\theta_V(k)\vert^2 = V \int dl\; \nu(l) \vert\theta^{(l)}_V(k)\vert^2,
\end{equation}
where $\vert\theta^{(l)}_V(k)\vert^2$ is the contribution from a domain of size $l$.  For the spectrum we then get
\begin{equation}\label{eqn:calcspectrum}
\delta^2_\theta = k^3 \int dl\; \nu(l) \vert\theta^{(l)}_V(k)\vert^2.
\end{equation}
We thus need to calculated the number $\nu(l)$ of regions and the contribution $\vert\theta^{(l)}_V(k)\vert^2$ from one such region. 

\begin{table}
\caption{The critical exponent $\eta$. }. \label{tab:eta}
\begin{center}
\begin{tabular}{clcc}
\hline
\hline
 & System & $\eta$ & \\
\hline
 & Mean-field & 0 & \\
 & \emph{3D theory} & & \\
 & n=1 (Ising) & 0.04 & \\
 & n=2 ($xy$-model) & 0.04 & \\
 & n=3 (Heisenberg) & 0.04 & \\
 & \emph{Experiment} & & \\
 & 3D n=1 & 0.03 -- 0.06 \\
\hline
\end{tabular}
\end{center}
This table has been adapted from \cite{chaikin}. $n$ is the dimension of the order parameter.
\end{table}

\section{The number of regions}
The leading behavior of $\nu(l)$ can be derived from the scale invariance of the system at the transition. We have defined $\nu(l)$ such that
\begin{equation}
V\nu(l)dl
\end{equation}
is the number of regions with sizes between $l$ and $l+dl$ in the volume $V$.  If we scale the whole system by $\lambda>0$ then the above expression gives the number of regions with sizes between $\lambda l$ and $\lambda l + \lambda dl$ in the volume $\lambda^d V$.  We can define the scaled number density $\nu_\lambda(l)$ by
\begin{equation}\label{eqn:rescale}
\nu_\lambda(l) = \frac{1}{\lambda^d}\nu(l/\lambda)\frac{1}{\lambda}.
\end{equation}
The last factor of $\lambda$ accounts for the change in $dl$.  Scale invariance\footnote{It is helpful here to think of the two steps that make up a renormalization group transformation.  Let us assume we devide the system into small cells of size $a$.  The first step of a renormalization group transformation would then consist of averaging neighboring cells.  This step does not change $\nu(l)$ for length scales larger than $a$.  The second step,  i.e. the rescaling,  changes $\nu(l)$ in the way given by equation (\ref{eqn:rescale}) and produces an equivalent result to the one we started out with.  This is what equation (\ref{eqn:scale free}) says.} then implies
\begin{equation}\label{eqn:scale free}
\nu_\lambda(l) = \nu(l).
\end{equation}
This gives the following scaling behavior for $\nu(l)$:
\begin{equation}\label{eqn:scaling}
\nu(l) \sim l^{-d-1}
\end{equation}
This is not the whole story though.  Because of the presence of a small length scale $a$ there is perfect correlation of the order parameter for length scales smaller than $a$.  For length scales larger than $a$ this discreteness shows up as a small change in the exponent of the power law behavior of the correlation function.  Instead of $G(r) \sim r^{-d+2}$ we have 
\begin{equation}\label{eqn:correlation}
G(r) \sim r^{-d +2 -\eta}.
\end{equation}
Typical values for $\eta$ are given in table \ref{tab:eta}.  Note how very little variation there is in the value of $\eta$ and how close it is to the observed value of the tilt of the spectrum of the cosmic microwave background.  We see that there is slightly \emph{more} correlation at smaller distances because of the existence of the length scale $a$.  We can repeat this argument for the number $\nu(l)$ of regions with size $l$.  For length scales smaller than $a$ the number of regions drops to zero.  The effect of the small length scale $a$ is thus to have slightly \emph{fewer} domains of small size.  The observations about the correlation function $G(r)$ and the number of regions $\nu(l)$ are consistent because it takes more regions of differing orientations to destroy correlation.  More correlation at smaller length scales implies fewer regions of small size.  In appendix \ref{app:nu} we argue that in fact
\begin{equation}\label{eqn:nuofl}
\nu(l) \sim l^{-d-1+\eta}.
\end{equation}

\section{The contribution from one region}
Next we have to calculate the contribution to the spectrum from one region of size $l$: 
\begin{equation}
\vert\theta^{(l)}(k)\vert^2
\end{equation}
We assume that the value of the order parameter $\theta$ is constant inside the region.  If one thinks of a spin model then the new ground state might be all spins down.  The constant value that we are talking about here is the value that corresponds to the spins being up.  In particular this implies that the constant does not depend on $l$.  Using Parseval's identity we obtain for $k<1/l$ that
\begin{equation}
l^3 \vert\theta^{(l)}(x)\vert^2 = l^3\cdot \text{const.} = \frac{1}{l^3}\vert\theta^{(l)}(k)\vert^2
\end{equation}
or
\begin{equation}\label{eqn:single}
\vert\theta^{(l)}(k)\vert^2 \sim l^{6}.
\end{equation}
The factor of $l^{-3}$ on the right hand side of the equation arises because the support of the Fourier transform will have size $1/l$ if the support of the function has size $l$.  To summarize
\begin{equation}\label{eqn:phicontribution}
\vert\theta^{(l)}(k)\vert^2 \sim \left\{ \begin{array}{ll}%
  l^6 & k<1/l \\
  0   & \text{else}
  \end{array}\right.
\end{equation}

\section{The result}
We are now in a position to calculate the spectrum $\delta^2_\theta$ of the order parameter just before the transition: 
\begin{equation}
\delta^2_\theta \sim k^3 \int dl\; \nu(l) \vert\theta^{(l)}_V(k)\vert^2
\end{equation}
Using equations (\ref{eqn:nuofl}) and (\ref{eqn:phicontribution}) we obtain
\begin{eqnarray}
\delta^2_\theta & \sim & k^3 \int^{1/k}_0 dl \; l^{-4+\eta}\; l^6 \\
  & \sim & k^3 \left(\frac{1}{k}\right)^{3+\eta}\\
  & \sim & k^{-\eta}
\end{eqnarray}
If we identify $\theta$ with the metric perturbation $\Phi$ this is exactly the spectrum found in the cosmic microwave background.  

\section{Conclusions}
In this paper we have pointed out a curious fact.  An almost scale free spectrum with a slight tilt of $0.04$ appears in two very different places in nature.  One place is early universe cosmology and the other is an ordinary phase transition.  One can have two different attitudes towards this result.  Either one thinks that this is pure numerology or one takes this to be an indication that a phase transition was part of the early history of the universe.

Let us investigate the second option.  The most important consequence is that it replaces the big bang and a period of inflation with a phase transition.  This might look like too high a price to pay but it it is worth remembering that the theory of inflation is not without problems.  The phase transition saves us from having to introduce a new field and a potential that is chosen just so to produce the desired tilt.  It also saves us from having to contemplate the initial value problem and the multiverse problem that automatically come with the theory of inflation.  It is no longer the case that we have to ask ourselves how it is that we live in this universe if everything that could happen did happen an infinite number of times.

Our approach also solves the problems that inflation originally set out to solve:  the horizon problem,  the flatness problem,  and the magnetic monopole problem.  The horizon problem is easily addressed because the size $\xi_0$ of the correlated domain is just related to the speed with which the phase transition is traversed.  The flatness problem doesn't arise because the solid state like models that we are looking at here naturally favor flatness.  Monopoles are expected to appear at the boundary of the correlated domains of size $\xi_0$ and are thus not visible in our universe.

The reason for the existence of the anomalous dimension $\eta$ is the presence of a small length scale.  In solid state physics this is the atomic spacing; here it implies the existence of a new fundamental small length scale that we have not yet observed in any other way.  The tilt of the cosmic microwave background is thus an indication that our world is discrete or at least possesses a further fundamental small length scale.  It is worth pointing out how robust the value of $\eta$ is.  It varies very little over different dimensions and systems (see table \ref{tab:eta}). 

Another important consequence of this result is what it implies for the search of a theory of quantum gravity.  The fact that we see the results of a phase transition in the cosmic microwave background shows that gravity has to be emergent.  It implies in particular that quantum gravity is not the quantization of gravity just like water molecules are not obtained by quantizing water waves.  Many attempts at a quantum theory of gravity deal with quantum space times  that are superpositions of Planck scale sized versions of space time \cite{loops,foams,cdt,causalsets}.  This result implies that quantum gravity is not of this kind.  

In this paper we have suggested to identify the order parameter with the gravitational potential.  This is what we have advocated in \cite{dreyer}.  The order parameter that we have looked at here is of course far too simple.  It is to be expected that the parameter that is describing our world is far more complicated than a scalar.  What we do expect though is that gravity is the effect of the non-perturbative structure of the vacuum.  It is our conviction that if one proposes such a drastic revision of the basic building blocks of physics some things should be simple.  This result is this simple thing.  Instead of a theory that leads us to reconsider the kind of questions science can answer we are just dealing with a simple phase transition.  

\appendix
\section{Fractals}\label{sec:fractal}
Let $F$ be a subset of $\R^d$.  For every $l>0$ let $N(l)$ be the number of spheres of radius $l$ that are needed to cover $F$.  When we let $l$ go to zero $N(l)$ typically diverges.  The way $N(l)$ diverges allows us to define the dimension $D$ of $F$.  We set
\begin{equation}
N(l) \sim l^{-D}.
\end{equation}
It turns out that there are sets $F$ for which this number $D$ is not an integer.  We will call $D$ the fractal dimension of the set $F$ (see \cite{fractal} for more details).  

Another way to characterize the set $F$ is through the correlation function $G(r)$ corresponding to $F$.  Let $\chi_F$ be the characteristic function of $F$:
\begin{equation}
\chi_F(x) = \left\{
  \begin{array}{l l}
    1 & \quad \text{if }x\in F\\
    0 & \quad \text{otherwise}
  \end{array} \right.
\end{equation}
Then we define 
\begin{equation}
G(x) = \frac{1}{V} \int_Vd^3y\; \chi_F(y+x) \chi_F(y).
\end{equation}
It turns out that the small scale behavior of the correlation function of $F$ is related to the fractal dimension of $F$.  If the correlation function behaves like
\begin{equation}
G(r) \sim r^{-\alpha}
\end{equation}
then the fractal dimension is
\begin{equation}
D = d - \alpha.
\end{equation}
This connection provides a convenient way to calculate the fractal dimension of a given set $F$.

\section{From $N(l)$ to $\nu(l)$.}\label{app:nuofl}\label{app:calc}
\begin{figure}[th]
\begin{center}
\includegraphics[width=8cm]{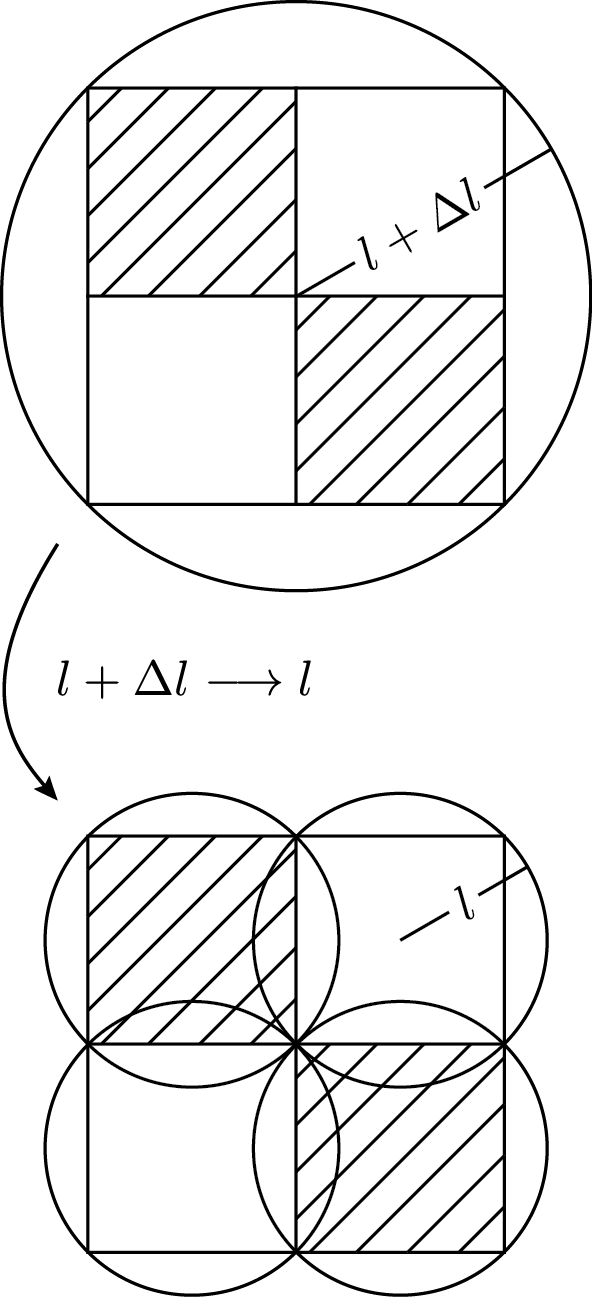}
\caption{The way to get the number $\nu(l)$ of regions of size $l$ from $N(l)$.}\label{fig:regions}
\end{center}
\end{figure}

For some region in $F\subset\R^d$ let $N(l)$ be the number of spheres of radius $l$ that are needed to cover $F$.  The number that we are looking for is just the number of regions that have size $l$.  The way to get this number is to see how $N(l)$ changes when we change $l$.  Let us say we go from $l+\Delta l$ to $l$.  If we have one solid region of size $l+\Delta l$ then we should have
\begin{equation}
N(l) = \left(\frac{l+\Delta l}{l}\right)^d N(l+\Delta l).
\end{equation}
The difference of the two sides of this equation is a measure for the number $\nu(l)$ of regions of size $l$ (see the figure \ref{fig:regions}).  Thus 
\begin{eqnarray}
\nu(l)\Delta l & = & N(l) + \Delta l N^\prime(l) + d \frac{\Delta l}{l} N(l) + O(\Delta l^2) - N(l)\\
  & = &  \Delta l \left( N^\prime(l)  + \frac{d}{l} N(l) \right)
\end{eqnarray}
Both terms go like $N(l)/l$.

\section{Calculating $\nu(l)$}\label{app:nu}
Let us start with the correlation function of equation (\ref{eqn:correlation}).  It corresponds to a fractal of dimension (see \ref{sec:fractal}) 
\begin{eqnarray}
D & = & d - (d-2+\eta) \\
  & = & 2 - \eta
\end{eqnarray}
The number of spheres of size $l$ that are needed to cover this fractal is given by
\begin{equation}
N_{\text{f}}(l) \sim l^{-2+\eta}.
\end{equation}
The number $\nu_{\text{f}}(l)\Delta l$ of regions with sizes between $l$ and $l+\Delta l$.  This number is proportional to
\begin{equation}
\Delta l\left( D\frac{N_{\text{f}}(l)}{l} + N_{\text{f}}^\prime(l)\right)
\end{equation}
(see appendix \ref{app:calc} for the details).  We just care about the power of $l$ which is 
\begin{equation}\label{eqn:refinedscaling}
l^{-3+\eta}.
\end{equation}
We now have to connect the number $\nu_{\text{f}}$ of regions of size $l$ that cover the fractal to the total number $\nu(l)$ of regions that flip.  If $\nu_u(l)$ and $\nu_d(l)$ are the number of domains of up and down spins of size $l$ then we have shown that
\begin{equation}
\Delta\nu(l) = \nu_u(l) - \nu_d(l) \sim l^{-3+\eta}.
\end{equation}
Let $p$ be the fraction of regions contributing to the fractal,  i.e.
\begin{equation}
p\nu_u(l) = \Delta\nu(l).
\end{equation}
The question is whether $p$ depends on $l$.  A look at equation (\ref{eqn:scaling}) shows that 
\begin{equation}
p\sim l^{d-2}.
\end{equation}
Thus
\begin{equation}\label{eqn:nrdomains}
  \nu(l) \sim l^{-d-1+\eta}.
\end{equation}

\begin{acknowledgments}  I want to thank FQXi,  the Foundational Questions Institute,  for financial support, Joao Magueijo,  Giovanni Amelino-Camelia,  Michele Arzano,  Giovanni Palmisano,  Mary Engleheart,  Seth Lloyd,  Lorenzo Sindoni,  and Fotini Markopoulou for comments on earlier versions of this paper,  and Mike Lazaridis for his vision and his support of fundamental science.
\end{acknowledgments}

\end{document}